\documentstyle[11pt]{article}
\normalbaselines
\begin{document}
\bibliographystyle{unsrt}
September 2, 1992\hfill
\def\ra{\rightarrow}
\def\al{\alpha}
\def\2pi{1\over 2\pi i}
\def\q{q-q^{-1}}
\def\.{\mathaccent"005F}
\def\={\mathaccent"0016}
\def\^{\mathaccent"007E}
\def\~{\tilde}
\def\newline{\hfil\break}
\def\sq2{\sqrt{2}}
\def\gam{\gamma}
\def\la{\lambda}
\def\ra{\rightarrow}
\def\va{\varphi}
\def\pa{\partial}
\def\sqk2{\sqrt{2(k+2}}
\def\sqk{\sqrt{k}}
\def\sqs{\sqrt{2\over k}}
\def\ps{\psi}
\def\psd{\psi^{\dagger}_1(z)}
\def\psdl{\psi^{\dagger}_\ell(z)}
\def\be{\begin{equation}}
\def\ee{\end{equation}}
\def\bea{\begin{eqnarray}}
\def\eea{\end{eqnarray}}
\vbox{\vspace{10mm}}
\vskip 1.5truecm
\begin{center}
{\LARGE \bf Deformation of the Wakimoto construction}\\[5mm]
A. Abada\,$^{*}$, A.H. Bougourzi\,$^{\dagger}$ and M.A.
El Gradechi\,$^{\dagger\,\spadesuit}$\\
[2mm]$^{*}${\it Institute for Theoretical Physics\\
State University of New York\\
Stony Brook, NY, 11794 USA\\
[2mm]$^{\dagger}$CRM,
Universit\'e de Montr\'eal\\
C.P. 6128-A\\
Montr\'eal, P.Q., Canada, H3C 3J7\\[2mm]
$^{\spadesuit}$Department of Mathematics\\
Concordia University\\
Montr\'eal, P.Q., Canada, H4B 1R6}\\[5mm] \end{center}

\begin{abstract}
We present the  extension of the Wakimoto construction to the
$su(2)_k$ quantum current algebra and its associated  $Z_k$
quantum parafermion algebra. This construction is achieved in
terms of various deformations of three classical free boson
fields.  We also  give the vertex operators corresponding to
the quantum spin-$j$ representation.
\end{abstract}

\section{Introduction}
It is well established now that  quantum groups and algebras play
an important role in both two-dimensional  integrable models of
quantum field theory and statistical mechanics, and  conformal field
theory \cite{Dri86,Jim86,Wor87,AlSi89,Dev89,FrRe92,BeLe92,Daval92}.
In particular,  many authors have realized the importance of the
infinite dimensional symmetries associated to quantum affine
algebras  (QAA) in constraining the integrable models.  In fact,
one hopes to carry over the success of the classical affine
symmetries  in chiral conformal field theory to the quantum
case.  In the classical case,  it has been realized through
many contributions that the Feigin-Fuchs construction (FFC)
\cite{FeFu82} is  convenient for  the technical resolution of
conformal field theory, that is, it facilitates the study and
the computation of several relevant physical quantities such as
the correlation functions, the BRST-like cohomology structure,
the irreducible characters and the fusion rules.  This is
because  the FFC consists in representing the integrable
models in terms of  free fields whose properties are well
known and  simple. Therefore, the extension of the FFC to the
quantum case  and in particular to the QAA is highly desirable
\cite{FrRe92,BeLe92,BaBe92}.

QAA were first introduced in Refs.  \cite{Dri85,Jim85}.
 After that, the  first FFC of the level one simply laced QAA
was  achieved by   Frenkel and Jing \cite{FrJi88}.  Then the
FFC of the spin-1/2 representation of the $sl(2)_1$ QAA
followed in Ref. \cite{Jimal92}. The FFC of the classical
$su(2)_k$ affine algebra was first obtained by Wakimoto
\cite{Wak86} in terms of a pair of ghost fields and a free
boson  field. After that, other   equivalent Wakimoto
constructions in terms of three free boson fields were
considered   \cite{ItKa90,Geral90,Bou92}.  This then led to
the FFC of the $Z_k$ parafermion theory
\cite{Neme89,Jayal90,GrHe90,Ahnal91}.

The purpose of this paper is to extend the Wakimoto construction
to the QAA $su(2)_k$. This paper is organized as follows. In
section 2, we review  the  $su(2)_k$ QAA in  the context of
conformal field theory. More specifically, we define the
operator product expansions satisfied by the currents
generating the $su(2)_k$ QAA, which is henceforth referred to
as the $su(2)_k$ quantum  current algebra (QCA). In section 3,
we first review the Frenkel-Jing construction of the $su(2)_1$
QCA, which requires different $q$-deformations of a single free
boson field. Then we use it {\it \`a la} Zamolodchikov and
Fateev  \cite{ZaFa85} to obtain the Wakimoto construction
of the $su(2)_k$ QCA. In this case, various $q$-deformations
of the three classical free boson fields are necessary.
As a by-product, the currents generating the $Z_k$ quantum
parafermion algebra are defined and their FFC is given. In
section 4, we derive the vertex (intertwining) operators
corresponding to the spin-$j$ representations of the $su(2)_k$
QCA and its associated $Z_k$ quantum parafermion algebra.

\section{ The $su(2)_k$ quantum current algebra}

The  $su(2)_k$ QAA is generated by the operators \{$E^\pm_n$,
$H_n$,  $\gamma^\pm$, $n\in {\bf Z}$\} as follows:
\bea
& &\gamma^{\pm} \in \quad{\rm the\: \:center\: \:of\: the\:
\:algebra},\\
& &{[H_n,H_m]} =  \delta_{n+m,0}{[2n]\over 2n}
{\gamma^{kn}-\gamma^{-kn}\over q-q^{-1}},\qquad n\neq 0,\\
& &{[H_0, H_m]}=  0,\\ & &{[H_n,E^{\pm}_m]}=
\pm\sqrt{2}{\gamma^{\mp |n|k/2}[2n]\over 2n}
E^\pm_{n+m}, \qquad n\neq 0,\\
& &{[H_0, E^\pm_m]} =  \pm\sq2 E^\pm_m,\\& &{[E^+_n,E^-_m]} =
{\gamma^{k(n-m)/2}\Psi_{n+m}-\gamma^{k(m-n)/2}
\Phi_{n+m}\over q-q^{-1}},\\
& &E^\pm_{n+1}E^\pm_m-q^{\pm 2}E^\pm_mE^\pm_{n+1}=
q^{\pm 2}E^\pm_nE^\pm_{m+1}-E^\pm_{m+1}E^\pm_n,
\eea
where $n, m\in {\bf Z}$, $[n]=(q^n-q^{-n})/q-q^{-1})$, $q$ is
the deformation parameter, and $\Psi_n$ and $\Phi_n$  are
respectively the modes of the fields $\Psi(z)$ and
$\Phi(z)$ ($z$ is a complex variable). These fields are
defined  by
\bea
& &\Psi(z)=\sum_{n\geq 0}\Psi_nz^{-n}=q^{\sqrt{2}H_0}
\exp[\sqrt{2}(\q)\sum_{n>0}H_nz^{-n}],\\
& &\Phi(z)=\sum_{n\leq 0}\Phi_nz^{-n}=q^{-\sqrt{2}H_0}
\exp[-\sqrt{2}(\q)\sum_{n<0}H_nz^{-n}].
\eea
Note that the algebra (1)-(7) coincides with the usual
$su(2)_k$ affine algebra \cite{Kac85} in the limit $q\ra 1$.

In order to derive the extension of the Wakimoto construction
to the above algebra it is convenient to express the relations
(2)-(7) as a QCA. This is a standard procedure in the context
of conformal field theory.  This QCA  is generated by the
currents  $E^\pm(z)$ and $H(z)$ (with $\gamma$ being replaced
by its eigenvalue $q$ \cite{FrJi88,Jimal92}), which are defined
as the generating functions of $E^\pm_n$ and $H_n$ respectively,
i.e.,
\bea
E^\pm(z)&=&\sum_{n=-\infty}^{+\infty}E^\pm_nz^{-n-1},\\
 H(z)&=&\sum_{n=-\infty}^{+\infty}H_nz^{-n-1}.
 \eea
The relations (2)-(7) are then equivalent to the following QCA,
which reads as operator product expansions (OPE) among
$E^\pm(z)$  and  $H(z)$:
\bea
\!\!\!\!\!\!\!\!\!\!\!\!&&H(z)H(w)=
\sum_{n>0}{[2n][nk]\over 2n}z^{-n+1}w^{n+1}+{\rm regular},\\
\!\!\!\!\!\!\!\!\!\!\!\!&&H(z)E^{\pm}(w)=\pm{\sqrt{2}\over
z}\left[1+ \sum_{n>0}{q^{\mp nk/2}[2n]\over 2n}z^{-n}w^n\right]
E^\pm(w) +{\rm regular},\\
 \!\!\!\!\!\!\!\!\!\!\!\!&&E^{\pm}(w)H(z)=\mp\sqrt{2}
\sum_{n>0} {q^{\mp nk/2}[2n]\over 2n}z^{n-1}w^{-n}E^\pm(w)+
{\rm regular},\\
\!\!\!\!\!\!\!\!\!\!\!\!&&E^+(z)E^-(w)={1
\over w(\q)}\left[{\Psi(wq^{k/2})\over z-wq^k}-
{\Phi(wq^{-k/2})\over z-wq^{-k}}\right]+{\rm regular},\qquad
|w|<\{|zq^k|,|zq^{-k}|\},\\
\!\!\!\!\!\!\!\!\!\!\!\!&&E^-(w)E^+(z)={1
\over w(\q)}\left[{\Psi(wq^{k/2})\over z-wq^k}-
{\Phi(wq^{-k/2})\over z-wq^{-k}}\right]+{\rm regular},
\qquad |z|<\{|wq^k| ,|wq^{-k}|\},\\
\!\!\!\!\!\!\!\!\!\!\!\!&&(z-w q^{\pm 2})E^{\pm}
(z)E^{\pm}(w)=(z q^{\pm 2}-w)
E^{\pm}(w) E^{\pm}(z),
\eea
where ``regular" refers to terms that are non-singular in
the  limit $z,w\ra 0$ in  (12)-(14) and  $z\ra wq^{\pm k}$
in (15) and (16) respectively. Note that in (15) and (16)
we have used  the identification $\sum_{n\geq 0}z^n\equiv
(1-z)^{-1}$ with $|z|<1$. This identification will also be
understood in the subsequent treatment, especially when
defining the contours of the Cauchy integrals.

\section{The Wakimoto construction of the $su(2)_k$ QCA}

As our construction relies on that of Frenkel-Jing in the
case of  $su(2)_1$ QCA \cite{FrJi88}, let us  first review
the latter in the  language of conformal field theory. It
is given by
\bea
H(z)&=&i\partial\xi^0(z),\\
E^\pm(z)&=&\exp[\pm i\sq2\xi^\pm(z)].
\eea
Here, $\xi^0(z)$ and $\xi^\pm(z)$ are deformed free bosonic
fields that are the generating functions of the basic
oscillators $\al_n$, $n\neq 0$, and the operators momentum
$P$ and position $Q$, that is,
\bea
\xi^0(z)&=&Q-iP\ln{z}+i\sum_{n\neq 0}{\al_n
\over n}z^{-n},\\
\xi^\pm(z)&=&Q-iP\ln{z}+i\sum_{n<0}
{q^{\pm n/2}\over [n]}\al_nz^{-n}+i
\sum_{n>0}{q^{\mp n/2}\over [n]}\al_nz^{-n},
\eea
with
\bea
{[\al_n,\al_m]}&=&\delta_{n+m,0}{[2n][n]\over 2n},\\
{[Q,P]}&=&i.
\eea
According to the usual definition of the normal ordering of
$\al_n$, $P$ and $Q$ as given in Ref. \cite{GoOl86}, the
relations (20)-(23) lead to the following two-point
correlation functions:
\bea
\langle \xi^\pm(z)\xi^\mp(w)\rangle&=&-{1\over 2}
\ln(z-wq)(z-wq^{-1}),\\
\langle \xi^\pm (z)\xi^\pm (w)\rangle&=&
-{1\over 2}\ln(z-w)(z-wq^{\mp 2}),\\
\langle \pa\xi^0(z)\xi^\pm(w)\rangle&=&
-z^{-1}\left[1+{1\over 2(\q)}\ln({z-wq^{-2\mp 1/2}\over
z-wq^{2\mp 1/2}})\right].
\eea
It can readily be seen through (24)-(26) that the $su(2)_1$
QCA (12)-(17) is indeed satisfied.

Let us  now make use {\it \`a la} Zamolodchikov  and Fateev
\cite{ZaFa85} of the Frenkel-Jing construction of $su(2)_1$ QCA
given in (18) and (19) to extend  the
Wakimoto construction to the $su(2)_k$ QCA (12)-(17).

The generalization of (18) so that $H(z)$ satisfies (12)
reads
\be
H(z)=i\sqk\partial\xi^0(z),
\ee
where now the deformed bosonic field $\xi^0(z)$ is defined by
\be
\xi^0(z)=Q-iP\ln{z}+i\sum_{n\neq 0}{\al_n\over n}z^{-n},
\ee
with
\bea
{[\al_n,\al_m]}&=&\delta_{n+m,0}{[2n][nk]\over 2nk},\\
{[Q,P]}&=&i.
\eea
Furthermore,  the natural generalization of $E^\pm(z)$ so
that (13) and (14) are satisfied reads
\be
E^\pm(z)=\exp\left[\pm i\sqs\xi^\pm(z)\right],
\ee
with the deformed free bosonic fields $\xi^\pm(z)$ being
\be
\xi^\pm(z)=Q-iP\ln{z}+ik\sum_{n<0}{q^{\pm nk/2}
\over [nk]}
 \al_nz^{-n}+ik\sum_{n>0}{q^{\mp nk/2}\over
[nk]}\al_nz^{-n}.
\ee
However, $E^\pm(z)$ as given  in (31) do not  satisfy
the relations (15)-(17)  of the $su(2)_k$ QCA. Instead,
they have the OPE:
\bea
E^\pm (z)E^\mp (w)&=& \exp \left[{2\over k}\langle
\xi^\pm(z)\xi^\mp(w)\rangle\right]:
E^\pm(z)E^\mp (w):,\\
E^\pm (z)E^\pm (w)&=& \exp \left[-{2\over k}\langle
\xi^\pm(z)\xi^\pm(w)\rangle\right]:E^\pm(z)E^\pm (w):,
\eea
where
\bea
\langle \xi^\pm(z)\xi^\mp(w)\rangle
&=&-\ln z+{k\over 2}\sum_{n>0}
{[2n]\over n[nk]}z^{-n}w^n,\\
\langle \xi^\pm(z)\xi^\pm(w)\rangle&=&
-\ln z+{k\over 2}\sum_{n>0}{q^{\mp nk}[2n]\over
n[nk]}z^{-n}w^n.
\eea
In (33) and (34), the symbol :: denotes the normal ordering
with respect to the bosonic operators $\al_n$, $P$ and
$Q$ \cite{GoOl86}. Clearly, the currents $E^\pm(z)$ as given
in (31) must be further  corrected by including  additional
deformed  bosonic fields in order to satisfy (15)-(17).
To this end, let $\va_1(z)$ and $\va_2(z)$ be two new deformed
bosonic fields with the expansions
\bea
\va_1(z)&=&Q_1-iP_1\ln z+i\sum_{n\neq 0}
{\beta_n\over n}z^{-n},\\
\va_2(z)&=&Q_2-iP_2\ln z+i\sum_{n\neq 0}
{\la_n\over n}z^{-n},
\eea
and
\bea
{[\beta_n, \beta_m]}&=&n\delta_{n+m,0}I(n),\\
{[\la_n,\la_m]}&=&-n\delta_{n+m}J(n),\\
{[Q_1,P_1]}&=&i,\\
{[Q_2,P_2]}&=&-i,
\eea
where $I(n)$ and $J(n)$ are  given by
\bea
I(n)&=&{k^2[n][n(k+2)/2]\over (k+2) [nk][nk/2]},\\
J(n)&=&{k\over 4}{[n(k+2)]+[2n]-[nk]\over [nk]}.
\eea
Note that all the remaining commutation relations are
trivial, $\va_2(z)$ has a ``time-like" signature, and in
the limit $q\ra 1$, $I(n), J(n)\ra 1$ as expected.
The currents $E^{\pm}(z)$ that fully satisfy the $su(2)_k$
QCA   (12) through (17) read  then  as follows:
\be
E^{\pm}(z)={\exp[\pm i\sqs\xi^{\pm}(z)]\over z(\q)}
\{\exp[iX^\pm(z,q)]-\exp[iX^\pm(z, q^{-1})]\},
\ee
where
\be
X^\pm(z,q)=\pm\sqs\va_2(zq^{k/2})+\sqrt{k+2\over 2k^2}
[\va_1(zq^k)-\va_1(z)].
\ee
It can easily be  checked that in the limit $q\ra 1$  the
correct classical expressions of the currents  $E^\pm(z)$
are recovered, namely, \cite{Bou92}
\be
E^\pm(z)=i\sqk\left[\pm {\pa\va_2(z)\over\sq2}+
\sqrt{k+2\over 2k}
\pa\va_1(z)\right]\exp\left\{\pm i\sqs[\va_2(z)+\xi(z)]\right\},
\ee
where $\xi(z)$ is the same classical limit of both
$\xi^\pm(z)$. For illustration, let us show  that  $E^\pm(z)$
satisfy the relations (15) and (16).  They have the  OPE
\bea
E^+(z)E^-(w)&=&\Theta(z,w,q)+{\rm regular},\qquad
|w|<\{|z q^k|,  |zq^{-k}|\},\\
E^-(w)E^+(z)&=&\Theta(z,w,q)+{\rm regular},\qquad
 |z|<\{|w q^k|,  |wq^{-k}|\},
\eea
with ``regular" referring  again to  non-singular terms
in the limit $z\ra wq^{\pm k}$, and $\Theta(z,w,q)$ being
defined by
\bea
\!\!\!\!\!\!\!\!\!\Theta(z,w,q)\!\!\!&=&-
{:\exp\{ i\sqs[\xi^+(z)\!-\!\xi^-(w)]\}:
\over zw(\q)^2} \{{q(z-w q^{-k-2})
\over z-wq^{-k}}:\exp[iX^+(z,q)\!-
\!iX^-(w,q^{-1})]:\\
&&+{z-w q^{k+2}\over  q(z-wq^k)}:\exp[iX^+(z,q^{-1})\!-
\!iX^-(w,q)]:\}.
\nonumber
\eea
Here the normal ordering is with respect to all the bosonic
oscillators. Using now standard rules of deriving the
commutation  relations from the OPE  \cite{GoOl86},
one finds that
\be
{[E^+_n, E^-_m]}={1\over (2\pi i)^2}\oint_{C_0}
dw w^m\oint_{C_1} dz z^n \Theta(z,w,q).
\ee
Because of the relations (48) and (49), the contours
$C_0$ and $C_1$ enclose respectively the origin point $w=0$
and the poles $z=wq^{\pm k}$.   The contribution of the
fields $\va_1(z)$ and $\va_2(z)$ cancel in a non-trivial way
after performing the  first integration in (51). In fact,
only  the contribution of the fields $\xi^\pm(z)$ survives as
expected, and because of the identities
\bea
\Psi(z)&\equiv &:\exp\left\{
i\sqs[\xi^+(zq^{k/2})-\xi^-(zq^{-k/2})]\right\}:,\\
\Phi(z)&\equiv &:\exp\left\{ i\sqs[\xi^+(zq^{-k/2})-\xi^-(zq^{k/2})]\right\}:,
\eea
it leads to the correct relations (15) and (16) or equivalently to the
relation (6). Finally, let us mention  that the
remaining relation (17) of the  $su(2)_k$  QCA is also
satisfied.

Let us now introduce the ``basic" $Z_k$ quantum parafermion
currents referred to as $\ps_1(z)$ and $\psd$.
As in the classical case, these currents are defined
by \cite{ZaFa85}
\bea
E^+(z)&=&\sqk\ps_1(z)\exp[ i\sqs\xi^+(z)],\\
E^-(z)&=&\sqk\psd\exp[ -i\sqs\xi^-(z)].
\eea
The FFC of $\ps_1(z)$ and $\psd$ follows from that of
$E^\pm(z)$ as given in (45) and (46). It reads as follows:
\bea
\ps_1(z)&=&{\exp[iX^+(z,q)]-\exp[iX^+(z, q^{-1})]
\over \sqk (\q)z},\\
\psd&=&{\exp[iX^-(z,q)]-\exp[iX^-(z, q^{-1})]
\over \sqk (\q)z}. \eea
The FFC of the remaining currents $\ps_\ell(z)$ and $\psdl$,
$\ell=2,\dots ,k$ generating the full quantum  $Z_k$
parafermion algebra can be derived from that of   $\ps_1(z)$
and $\psd$ through their OPE.

To recapitulate, the relations (27) and (45) describe the
Wakimoto construction  of the $su(2)_k$ QCA (12)-(17) in
terms of the deformed free bosonic fields $\xi^0(z)$,
$\xi^\pm(z)$, $\va_1(z)$ and $\va_2(z)$, which are given
in (28), (32), (37) and (38) respectively. The FFC of the
basic currents of the $Z_k$ quantum parafermion  algebra,
which are defined through (54) and (55), is given in (56)
and (57).

\section{Vertex realization of the spin-$j$ representation}

The spin-$j$ representation of the $su(2)_k$ QCA (12)-(14)
is described by the ``$q$-primary" fields $V^{j,m}(z)\equiv
V^m(z)$ ($m=-j,\dots,+j$)  and their ``$q$-descendants."
$V^{j}(z)$, which creates  the highest weight state from
the bosonic vacuum, is such that
\bea
{[E^+(z),V^{j}(w)]}&=&0,\\
{[H_n, V^j(w)]}&=&j\sq2 \left\{\delta_{n,0}+
\delta_{n>0}\gam^{nk/2}{[nk]\over nk}w^n
 +\delta_{n<0}\gam^{-nk/2}
{[nk]\over nk}w^n \right\}V^j(w).
\eea
The FFC of $V^j(z)$ that is consistent with (58) and (59) is
\be
V^j(z)=\exp\left\{ ij\sqs\left[\~\xi(z)+\~\phi_2(z)+
\sqrt{k\over k+2}\~\phi_1(z)\right]\right\},
\ee
where
\bea
\~\xi(z)&=&Q-iP\ln(-z)+2i\sum_{n>0}{q^{nk/2}
\over [2n]}\al_nz^{-n}+2i\sum_{n<0}{q^{-nk/2}
\over [2n]} \al_n z^{-n},\\
\~\phi_1(z)&=&Q_1-iP_1\ln(-z)+i\sum_{n>0}
{\beta_n\over nI(n)} z^{-n}+i\sum_{n<0}
{\beta_n\over nI(n)}z^{-n},\\
\~\phi_2(z)&=&Q_2-iP_2\ln(-z)+i\sum_{n>0}
{q^{nk/2}+q^{-nk/2}\over 2nJ(n)}\lambda_n
z^{-n}+i\sum_{n<0}{q^{nk/2}+q^{-nk/2}
\over 2nJ(n)}\lambda_n z^{-n}.
\eea
The FFC of the other fields $V^m(w)$ in the multiplet
are obtained from that of $V^j(w)$ through  the
relation
\be
V^m(w)=[E^-_0,\dots [E^-_0,V^j(w)]_{q^k}
\dots]_{q^k}.
\ee
In (64) there are $j-m$ deformed commutators, which  are defined by
\be
[E^-_0,V^j(w)]_{q^k}\equiv q^kE^-_0V^j(w)-
V^j(w)E^-_0.
\ee

Finally, let us note that the primary field $\chi^j(z)$
of the  $Z_k$  quantum parafermion algebra is again defined
as in the classical case, that is, through the relation
\be
V^j(z)=\chi^j(z)\exp\left[ ij\sqs\~\xi(z)\right].
\ee
This means that the FFC of $\chi^j(z)$ is given by
\be
\chi^j(z)=\exp\left\{ ij\sqs\left[\~\phi_2(z)+
\sqrt{k\over k+2}\~\phi_1(z)\right]\right\}.
\ee

While typing this paper we received a preprint by A.
Matsuo \cite{Mat92} discussing another deformation
of the Wakimoto realization.  However, he did not
discuss the FFC of the spin-$j$ representation and
the $Z_k$ quantum parafermion algebra introduced here.

\section{Acknowledgments}
This work was supported in part by funds provided by the
Natural Sciences and Engineering Research Council (NSERC)
of Canada, and the ``Fonds FCAR pour la formation de
chercheurs et l'aide \`a la recherche." We thank Professors
Y. Saint-Aubin,  V. Hussin and S.T. Ali for their
encouragement.

\end{document}